\documentclass[aip,apl,reprint]{revtex4-1}
\usepackage{graphicx}
\usepackage{dcolumn}
\usepackage{bm}
\usepackage{color}
\usepackage{amsmath} 
\graphicspath{{./figures/}}

\newcommand{\mathcommand}[3][0]{\newcommand{#2}[#1]{\ensuremath{#3}}}
\newcommand{\be}{\begin{equation}}
\newcommand{\ee}{\end{equation}}
\mathcommand[1]{\ket}{\left| #1 \right\rangle}  
\mathcommand[1]{\bra}{\left\langle #1 \right|}  
\mathcommand[2]{\ketbra}{
    \left| #1 \right\rangle\!\!\left\langle #2 \right|} 

\mathcommand[2]{\ts}{#1_{\text{#2}}}
\mathcommand{\te}{\text{e}}
\mathcommand{\thole}{\text{h}}
\mathcommand{\nodag}{{\phantom{\dag}}}

\mathcommand\gtwo{g^{(2)}(0)}

\newcommand{\TLfixA}[1]{#1}

\newcommand{\NII}{{
    National Institute of Informatics,
    2-1-2 Hitotsubashi, Chiyoda-ku,
    Tokyo 101-8430, Japan
}}

\newcommand{\Ginzton}{{
    Edward L. Ginzton Laboratory,
    Stanford University,
    Stanford, California 94305-4088, USA
}}

\newcommand{\Paderborn}{{
    Dept. of Physics,
    University of Paderborn,
    Warburger Str. 100, 33098
    Paderborn, Germany
}}
\newcommand{\HRL}{{
    Currently at HRL Laboratories, LLC, 3011 Malibu
Canyon Rd., Malibu, CA 90265.
}}

\begin{document}
\title{Photon Antibunching and Magnetospectroscopy of a Single Fluorine Donor in ZnSe}
\author{K. De Greve}
    \email[Electronic address: ]{kdegreve@stanford.edu}

\author{S. M. Clark}
\author{D. Sleiter}
\affiliation\Ginzton
\author{K. Sanaka}
\affiliation\Ginzton
\affiliation\NII
\author{T. D. Ladd}
\altaffiliation\HRL
\affiliation\Ginzton
\affiliation\NII

\author{M. Panfilova}
\affiliation\Paderborn
\author{A. Pawlis}
\affiliation\Ginzton
\affiliation\NII
\affiliation\Paderborn
\author{K. Lischka}
    \affiliation\Paderborn
\author{Y. Yamamoto}
    \affiliation\Ginzton
    \affiliation\NII

\begin{abstract}
We report on the optical investigation of single
    \TLfixA{electron spins bound to fluorine donor impurities}
in ZnSe.
    \TLfixA{Measurements of photon antibunching}
establish the presence of single, isolated optical emitters, and
magneto-optical studies are consistent with the presence of an
exciton bound to the spin-impurity complex. The isolation of this
single donor-bound exciton complex and its potential homogeneity
offer promising prospects for a scalable semiconductor qubit
    \TLfixA{with an optical interface}.
\end{abstract}

\pacs{78.55.Et, 
      78.67.-n,  
      03.67.-a}  


\maketitle



Schemes for quantum information processing and quantum communications rely on scalable, robust qubits.  In particular, there are many proposals that require fast, efficient, and homogenous single-photon sources~\cite{KLM2001,BB84,nielsenchuang} and still others that rely on the interaction between matter qubits and flying photonic qubits~\cite{divicenzo}.  The requisites for both types of schemes can be satisfied with semiconductor electron spins, which serve as single photon sources~\cite{charlieprl} or long lived quantum memories with an optical interface~\cite{Greilich2006, Clark2009, Press2010}.  However, optical schemes, particularly those based on entanglement, also require large numbers of homogenous photon emitters~\cite{czkm97, yls05, ctsl05, wv06, hybridnjp, susan}.  Electron spins in self-assembled QDs, unfortunately, suffer from large inhomogenities due to their natural size distribution.

Impurity-bound electrons in direct bandgap semiconductors, however, have relatively little inhomogeneous broadening~\cite{Lampert,
Haynes, Hopfield, merz, dean, kaoru_indistinguishable}, yet still possess strong optical transitions when binding an additional exciton~\cite{Karasyuk94a, merz, dean, kaoru_indistinguishable} and long ground state coherence times~\cite{Clark2009}.  

An electron bound to a single fluorine donor in ZnSe (F:ZnSe) may serve as a physical qubit with many potential advantages over previously researched qubits.  F:ZnSe is particularly appealing
because of its nuclear structure compared to III-V-based bound-exciton or quantum dot systems. Unlike III-V systems, isotopic purification of the ZnSe-host matrix to a nuclear-spin-0 background is possible,
eliminating magnetic noise from nuclear spin diffusion~\cite{Tyryshkin, desousa}.  Further, the F-impurity has a
nuclear spin of $1/2$ with 100\% abundance. Electron-nuclear spin
swapping schemes~\cite{benjamin_broker, Cakir2009} can be used, which, in combination with the spin-0 background of the
isotopically purified host matrix, could lead to an extremely
long-lived qubit.  Additionally, the applicability of standard microfabrication techniques~\cite{Pawlis2008,Panfilova2009} to ZnSe
makes the F:ZnSe system particularly scalable.

The F:ZnSe system has already shown promise as a scalable source of
single photons in Ref.~\onlinecite{kaoru_indistinguishable}.  However, this work did not demonstrate the
potential of the donor system as a future quantum memory.  Here, we
show both statistics for single photon emission, as well as the presence of a three-level optical $\Lambda$-system through magnetospectroscopy experiments. This introduces F:ZnSe as a valid candidate for use as a scalable qubit
with an optical interface.

    \begin{figure}[ht]
    \centering
    \includegraphics[width=8cm]{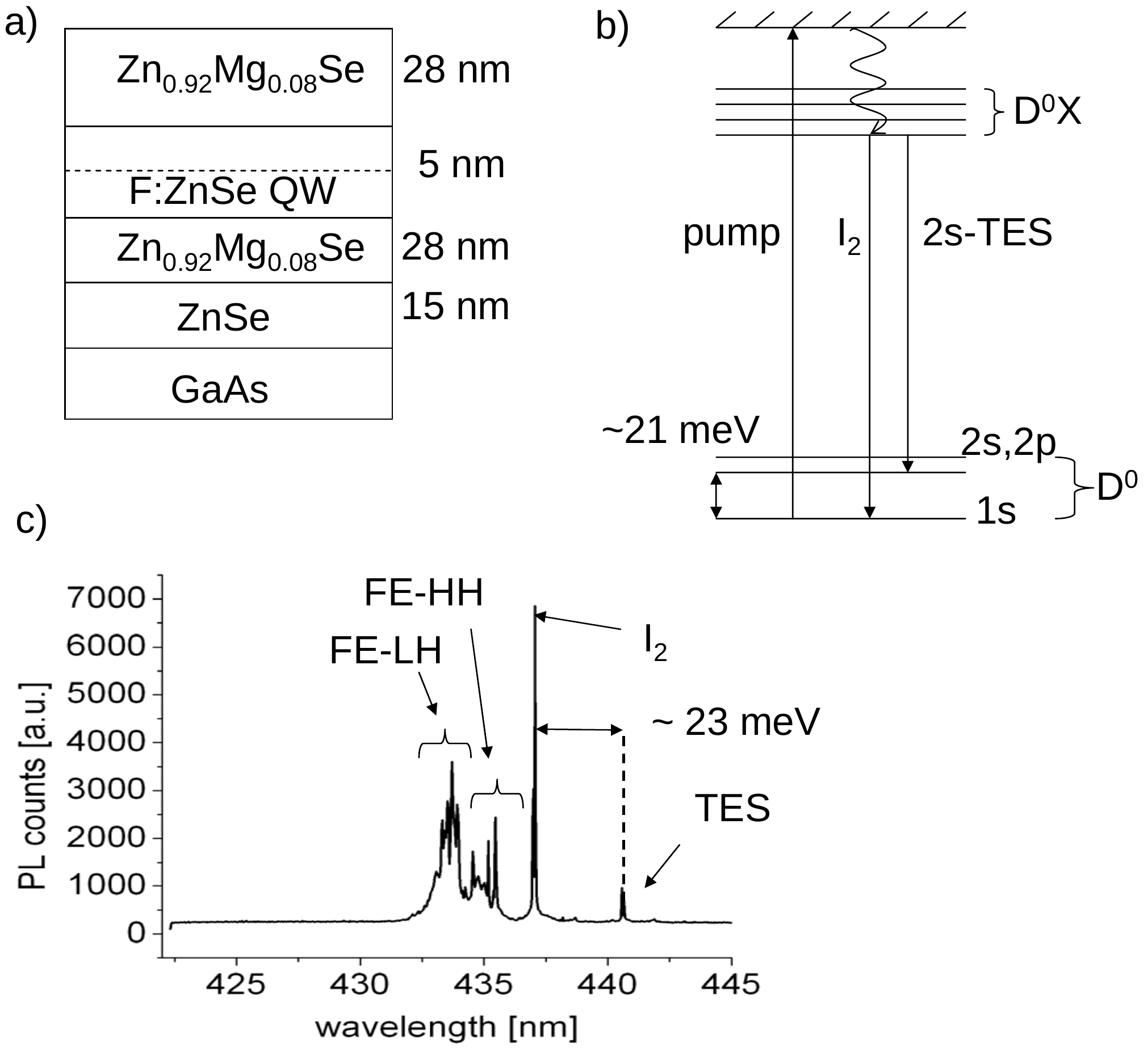}
    \caption[sample-structure and spectrum]{
        Sample structure and PL data.
        a) Schematic of sample structure: the dotted line indicates the location of the $\delta$-doping.
        b) level structure: D$^{0}$ represents the F-bound-electron manifold with its excited electron states, while D$^{0}$X represents the F-bound exciton manifold. Excitation
        occurs through above-band pumping, with fast non-radiative relaxation into the bound-exciton ground states. Radiative decay occurs into either the ground state (indicated by I$_{2}$) or
        excited states (TES).
        c) mesa-structure (100 nm diameter) PL. FE-HH and FE-LH represent the heavy and light hole free exciton emission, while I$_{2}$ and TES indicate the bound exciton decay into the respective electron ground and excited states.
        }
    \label{spectra}
    \end{figure}

The sample structure is illustrated in
    \TLfixA{Fig.~\ref{spectra}(a)}.
The samples that were studied all
    \TLfixA{consist}
of an MBE-grown, fluorine $\delta$-doped ZnSe quantum well (QW), ranging between 1 and 10 nm in thickness, sandwiched between
two ZnMgSe cladding layers (Mg content approximately 8\%).  GaAs-(001) substrates were used, with a thin, undoped ZnSe buffer layer defining a clean interface for the II-VI/III-V heteropitaxy.
The ZnMgSe cladding layers prevent carrier diffusion into the lower-bandgap GaAs
substrate and spectrally shift the background ZnMgSe emission from the F:ZnSe photoluminescence (PL). The
areal Fluorine impurity $\delta$-doping density was approximately
8 x $10^{9}$ cm$^{-2}$. $\delta$-doping was chosen both to ease the
isolation of individual impurities by reducing the total number of
dopants in the QW, as well as to locate the dopant
impurities in the center of the QW. Spatial isolation of individual emitters occurred through
mesa-etching, where selected parts of the Zn(Mg)Se-sandwich layer
were chemically removed in a K$_{2}$Cr$_{2}$O$_{7}$:HBr:H$_{2}$O
(1:130:250) etch mixture~\cite{Illing}. The nominal diameter of the mesas varied
from 50 \TLfixA{to} 400~nm, and the mesas were capped in SiO$_{2}$
for chemical stability.

PL characterization was
performed in a cold-finger, liquid-He cryostat at \TLfixA{10~K}, using a \TLfixA{408~nm} laser diode acting as
above-band pump source, similar to the experiments described
in Refs.~\onlinecite{Pawlis2006},\onlinecite{Kaorulasing}. The relevant transitions are schematically illustrated in Fig.~\ref{spectra}(b). A spectrum is shown in
    \TLfixA{Fig.~\ref{spectra}(c)}.
Note that several peaks are visible: I$_{2}$ and the two-electron satellites (TES)~\cite{dean} are related to the bound exciton decay, whereas FE-HH and FE-LH represent a continuum of heavy- and light hole free-exciton emission respectively. In contrast to the decay into the ground state (I$_{2}$), the TES are associated with decay into excited states of the bound electron. From a zeroth-order Bohr-model, these are expected to be redshifted from I$_{2}$ by about 21 meV, which is close to the
experimental value of 23 meV, hence establishing the
origin of the emission peaks as related to donor-bound exciton transitions. From separate, F-concentration dependent growth studies (not shown), we infer its origin as F-impurity bound exciton decay.

In order to verify whether a single peak corresponds to a single
emitter, photon correlation measurements \TLfixA{of
$g^{(2)}(\tau)$} were performed~\cite{HanburyBrown1958}. In these experiments, an aboveband pulsed laser excites the mesa of interest and the emitted photons are frequency-filtered using a grating and a slit.  This filtered light is sent to a beamsplitter with a detector at each output arm.  Coincidence clicks between the detectors are counted as a function of the delay, $\tau$, between those clicks.  After normalization by single detector clicks, the count rate, $g^{(2)}(\tau)$ gives insight to the nature of the light source via its photon statistics.  In such a measurement
the correlation function $g^{(2)}(\tau)$ for a perfect
single emitter \TLfixA{without a background}
    \TLfixA{approaches}
zero at zero delay ($\tau=0$)~\cite{mandelwolf,charlieprl}. A mode-locked laser
with repetition rate of 76~MHz was used for above-band pumping, and
silicon \TLfixA{avalanche-photodiode-based}
\TLfixA{single-photon-counting} modules were used with a time-interval analyzer for timing photon detection events.

       \begin{figure}[ht]
      \centering
      \includegraphics[width=8cm]{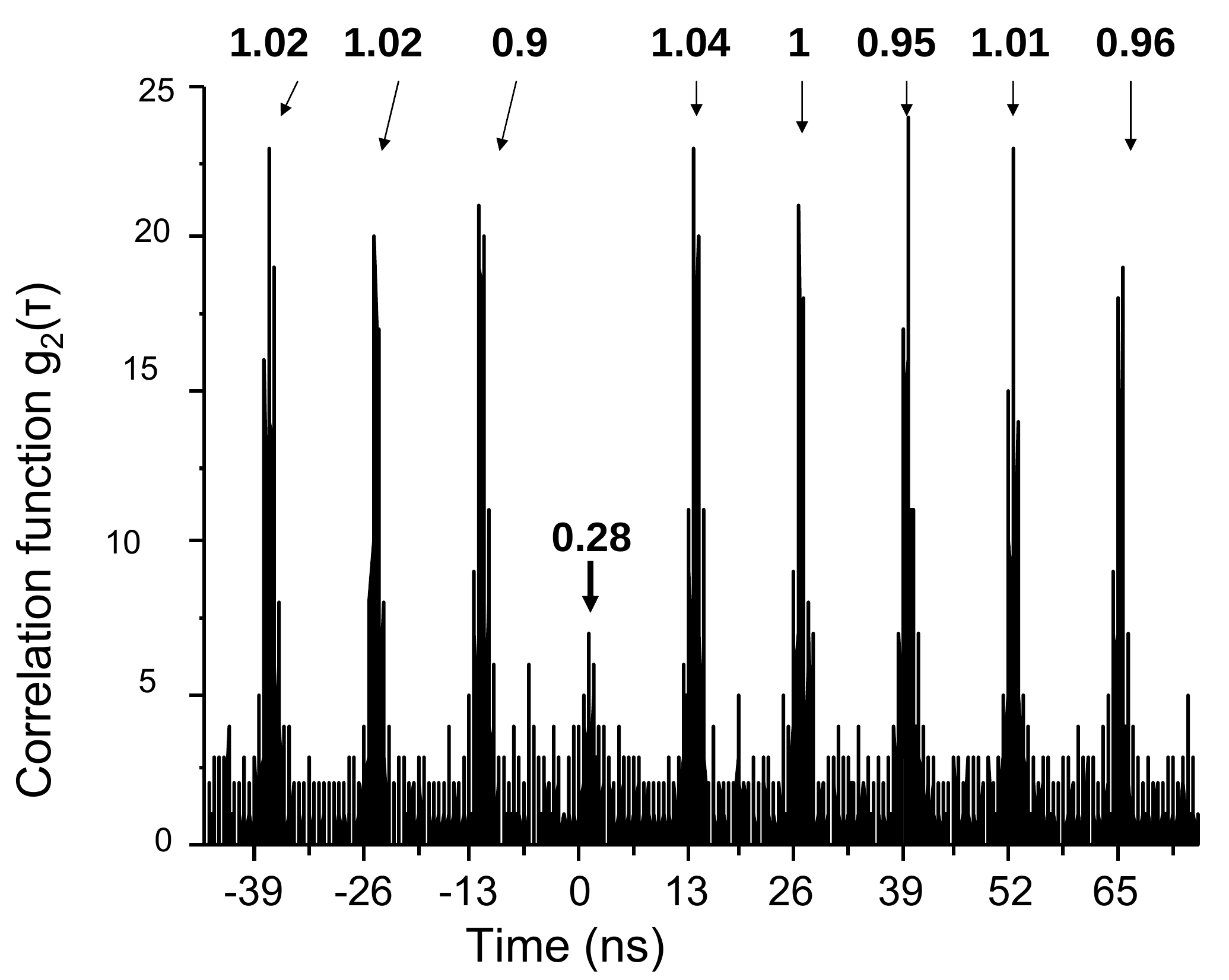}
      \caption[gtwo]{Non-normalized $g^{(2)}(\tau)$ correlation function measurement for a 100 nm mesa. Normalized peak values (1 ns integration window around the peaks) are indicated above the peaks . Note the \gtwo-dip, denoting antibunching, and the repetition rate of 13 ns or 76 MHz.}
      \label{gtwo}
      \end{figure}

    \TLfixA{The lowest \gtwo\ measured to date in these samples is
    0.28, as shown in Fig.~\ref{gtwo} for a mesa structure \TLfixA{with diameter} 100~nm, QW-thickness of 5 nm.}
    \TLfixA{This value confirms}
the presence of an individual emitter,
    \TLfixA{since it is less than the
    threshold value of 0.5 that excludes two or more photons being emitted~\cite{mandelwolf}.}
The origin of the 0.28 background is attributed to the presence of
nearby emitters, especially \TLfixA{from} the tail of the
free-exciton emission
peak.

\begin{figure}[ht]
   \centering
   \includegraphics[width=9cm]{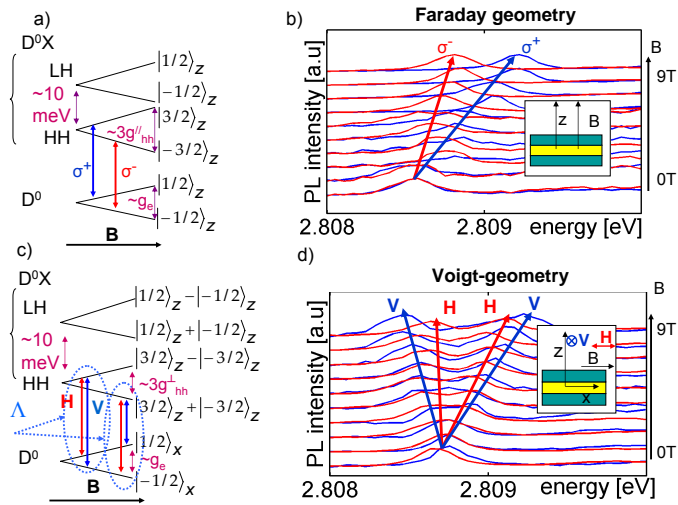}
   \caption[gtwo]{Magnetospectroscopy data for 200 nm diameter mesas. a) Energy spectrum in Faraday geometry
   (D$^{0}$X: bound exciton, D$^{0}$: bound electron, HH: heavy hole, LH: light hole); b) PL as function of magnetic field, in Faraday geometry (inset: sample and field alignment
   in Faraday geometry);
   c) Energy spectrum in Voigt geometry; note the two $\Lambda$-systems with distinct polarization selection rules; d)
   PL as function of magnetic field in Voigt geometry. (inset: sample and field alignment)
   }
   \label{magneto}
   \end{figure}

\TLfixA{While the measurement of $\gtwo$ establishes F:ZnSe as a
viable single-photon source, its use as an optically controllable qubit requires verification of the presence of a three-level $\Lambda$-system.
For this reason,}
magnetospectroscopy measurements were performed in a 10~T
superconducting magnet at \TLfixA{1.5~K}. A room-temperature reentrant bore window
allowed for a microscope objective (working distance: 38 mm, NA:
0.18) to be used for collection of the photoluminescence. Above-band
illumination \TLfixA{from a} CW, 408~nm GaN laser-diode was chosen
for the excitation of the sample, and the photoluminescence spectrum
was collected through a 750~mm spectrometer on a liquid-N$_2$-cooled
CCD-camera. Both Faraday and Voigt geometry data were obtained through the use of small mirrors inside the cryostat (We refer to the insets of Fig.~\ref{magneto}(b) and~(d) for the respective orientations).  


The expected energy spectrum as a function of applied magnetic field
is shown in
    \TLfixA{Figs.~\ref{magneto}(a) and (c)}
(See also Refs.~\onlinecite{merz}, \onlinecite{Karasyuk94a}, \onlinecite{Willmann}, \onlinecite{White1974} and \onlinecite{Dean1977} for a review on magnetospectroscopy of bound-exciton systems). Note that, due to
the QW and compressive strain in the ZnSe-layer, the degeneracy between the heavy
and light holes is lifted~\cite{Pawlis2006,Kaorulasing}.

In Faraday geometry, a two-fold split of the photoluminescence is
observed. This is consistent with the model of an exciton bound
to an electron-impurity complex.  
From the linesplit, we infer a difference in the out-of-plane heavy-hole and electron
$g$-factors ($\mid3g^{\parallel}_{\text{hh}} - g_\text{e}\mid$) of 0.8 ($\pm 0.2$).

The Voigt geometry data show a fourfold split, with linear polarization, again consistent with
\TLfixA{donor-bound-exciton} emission. From the linesplits, we infer
an electron $g$-factor of 1.2 ($\pm 0.2$), which can be compared with the value
of 1.36 for a donor-impurity bound electron as obtained by Wolverson \emph{et al.} through spin-flip
Raman scattering~\cite{Wolverson}. The heavy holes in the bound exciton complex are weakly coupled by the magnetic field, leading to an in-plane heavy-hole g-factor ($\mid3g^\perp_\text{hh}\mid$) of 0.2 ($\pm 0.2$). 
The Voigt data also establish the presence of two doubly-connected $\Lambda$-systems, as
illustrated in
    \TLfixA{Fig.~\ref{magneto}(c)},
establishing the F:ZnSe system as a candidate for use in several
proposed quantum information technology schemes~\cite{ctsl05,
hybridnjp, wv06, yls05, czkm97, duan04}. 



In conclusion, we isolated the emission of bound-excitons from a single F:ZnSe complex through mesa-etching and spectral filtering.  These methods establish F:ZnSe as a potentially homogeneous and efficient single photon source.  The energy splittings from the Faraday and Voigt magneto-photoluminescence data and the presence of TES emission firmly establish the presence of optically controllable electron spins bound to neutral donors in ZnSe.  The potential for nuclear purification of the ZnSe host matrix and fabrication ease makes F:ZnSe an attractive candidate for a long-lived, scalable quantum qubit.  Future work includes the incorporation of a microcavity~\cite{Kaorulasing} to enhance emission and enable efficient quantum networking.

This work was supported by NICT, MEXT, NIST 60NANB9D9170, NSF CCR-08 29694, University of Tokyo Special Coordination Funds for Promoting Science and Technology, and by the Japan Society for the Promotion of Science
(JSPS) through its Funding Program for World-Leading Innovative R$\&{}$D on
Science and Technology (FIRST Program). S.M.C was partially supported by the HP Fellowship Program through the Center for Integrated Systems, D.S. was partially supported by Matsushita through the Center for Integrated Systems, and M. P. and A. P. acknowledge the financial support in the DFG Graduiertenkolleg GRK-1464.




\begin{thebibliography}{37}
\expandafter\ifx\csname natexlab\endcsname\relax\def\natexlab#1{#1}\fi
\expandafter\ifx\csname bibnamefont\endcsname\relax
  \def\bibnamefont#1{#1}\fi
\expandafter\ifx\csname bibfnamefont\endcsname\relax
  \def\bibfnamefont#1{#1}\fi
\expandafter\ifx\csname citenamefont\endcsname\relax
  \def\citenamefont#1{#1}\fi
\expandafter\ifx\csname url\endcsname\relax
  \def\url#1{\texttt{#1}}\fi
\expandafter\ifx\csname urlprefix\endcsname\relax\def\urlprefix{URL }\fi
\providecommand{\bibinfo}[2]{#2}
\providecommand{\eprint}[2][]{\url{#2}}

\bibitem[{\citenamefont{Knill et~al.}(2001)\citenamefont{Knill, Laflamme, and
  Milburn}}]{KLM2001}
\bibinfo{author}{\bibfnamefont{E.}~\bibnamefont{Knill}},
  \bibinfo{author}{\bibfnamefont{R.}~\bibnamefont{Laflamme}}, \bibnamefont{and}
  \bibinfo{author}{\bibfnamefont{G.}~\bibnamefont{Milburn}},
  \bibinfo{journal}{Nature} \textbf{\bibinfo{volume}{409}}, \bibinfo{pages}{46}
  (\bibinfo{year}{2001}).

\bibitem[{\citenamefont{Bennett and Brassard}(1984)}]{BB84}
\bibinfo{author}{\bibfnamefont{C.}~\bibnamefont{Bennett}} \bibnamefont{and}
  \bibinfo{author}{\bibfnamefont{G.}~\bibnamefont{Brassard}},
  \bibinfo{journal}{Proceedings of the IEEE International Conference on
  Computer, Systems and Signal processing (IEEE press, NY, 1984)} p.
  \bibinfo{pages}{175} (\bibinfo{year}{1984}).

\bibitem[{\citenamefont{Nielsen and Chuang}(2000)}]{nielsenchuang}
\bibinfo{author}{\bibfnamefont{M.}~\bibnamefont{Nielsen}} \bibnamefont{and}
  \bibinfo{author}{\bibfnamefont{I.}~\bibnamefont{Chuang}},
  \emph{\bibinfo{title}{Quantum Computation and Quantum Information}}
  (\bibinfo{publisher}{Cambridge University Press}, \bibinfo{year}{2000}).

\bibitem[{\citenamefont{DiVicenzo}(2000)}]{divicenzo}
\bibinfo{author}{\bibfnamefont{D.}~\bibnamefont{DiVicenzo}},
  \bibinfo{journal}{Fortschr. Phys.} \textbf{\bibinfo{volume}{48}},
  \bibinfo{pages}{771} (\bibinfo{year}{2000}).

\bibitem[{\citenamefont{Santori et~al.}(2001)\citenamefont{Santori, Pelton,
  Solomon, Dale, and Yamamoto}}]{charlieprl}
\bibinfo{author}{\bibfnamefont{C.}~\bibnamefont{Santori}},
  \bibinfo{author}{\bibfnamefont{M.}~\bibnamefont{Pelton}},
  \bibinfo{author}{\bibfnamefont{G.}~\bibnamefont{Solomon}},
  \bibinfo{author}{\bibfnamefont{Y.}~\bibnamefont{Dale}}, \bibnamefont{and}
  \bibinfo{author}{\bibfnamefont{Y.}~\bibnamefont{Yamamoto}},
  \bibinfo{journal}{Phys. Rev. Lett.} \textbf{\bibinfo{volume}{86}},
  \bibinfo{pages}{1502} (\bibinfo{year}{2001}).

\bibitem[{\citenamefont{Greilich et~al.}(2006)\citenamefont{Greilich, Yakovlev,
  Shabaev, Efros, Yugova, Oulton, Stavarache, Reuter, Wieck, and
  Bayer}}]{Greilich2006}
\bibinfo{author}{\bibfnamefont{A.}~\bibnamefont{Greilich}},
  \bibinfo{author}{\bibfnamefont{D.~R.} \bibnamefont{Yakovlev}},
  \bibinfo{author}{\bibfnamefont{A.}~\bibnamefont{Shabaev}},
  \bibinfo{author}{\bibfnamefont{A.~L.} \bibnamefont{Efros}},
  \bibinfo{author}{\bibfnamefont{I.~A.} \bibnamefont{Yugova}},
  \bibinfo{author}{\bibfnamefont{R.}~\bibnamefont{Oulton}},
  \bibinfo{author}{\bibfnamefont{V.}~\bibnamefont{Stavarache}},
  \bibinfo{author}{\bibfnamefont{D.}~\bibnamefont{Reuter}},
  \bibinfo{author}{\bibfnamefont{A.}~\bibnamefont{Wieck}}, \bibnamefont{and}
  \bibinfo{author}{\bibfnamefont{M.}~\bibnamefont{Bayer}},
  \bibinfo{journal}{Science} \textbf{\bibinfo{volume}{313}},
  \bibinfo{pages}{341} (\bibinfo{year}{2006}).

\bibitem[{\citenamefont{Clark et~al.}(2009)\citenamefont{Clark, Fu, Zhang,
  Ladd, Stanley, and Yamamoto}}]{Clark2009}
\bibinfo{author}{\bibfnamefont{S.~M.} \bibnamefont{Clark}},
  \bibinfo{author}{\bibfnamefont{K.-M.~C.} \bibnamefont{Fu}},
  \bibinfo{author}{\bibfnamefont{Q.}~\bibnamefont{Zhang}},
  \bibinfo{author}{\bibfnamefont{T.~D.} \bibnamefont{Ladd}},
  \bibinfo{author}{\bibfnamefont{C.}~\bibnamefont{Stanley}}, \bibnamefont{and}
  \bibinfo{author}{\bibfnamefont{Y.}~\bibnamefont{Yamamoto}},
  \bibinfo{journal}{Phys. Rev. Lett.} \textbf{\bibinfo{volume}{102}},
  \bibinfo{pages}{247601} (\bibinfo{year}{2009}).

\bibitem[{\citenamefont{Press et~al.}(2010)\citenamefont{Press, De~Greve,
  McMahon, Ladd, Friess, Schneider, Kamp, Hofling, Forchel, and
  Yamamoto}}]{Press2010}
\bibinfo{author}{\bibfnamefont{D.}~\bibnamefont{Press}},
  \bibinfo{author}{\bibfnamefont{K.}~\bibnamefont{De~Greve}},
  \bibinfo{author}{\bibfnamefont{P.~L.} \bibnamefont{McMahon}},
  \bibinfo{author}{\bibfnamefont{T.~D.} \bibnamefont{Ladd}},
  \bibinfo{author}{\bibfnamefont{B.}~\bibnamefont{Friess}},
  \bibinfo{author}{\bibfnamefont{C.}~\bibnamefont{Schneider}},
  \bibinfo{author}{\bibfnamefont{M.}~\bibnamefont{Kamp}},
  \bibinfo{author}{\bibfnamefont{S.}~\bibnamefont{Hofling}},
  \bibinfo{author}{\bibfnamefont{A.}~\bibnamefont{Forchel}}, \bibnamefont{and}
  \bibinfo{author}{\bibfnamefont{Y.}~\bibnamefont{Yamamoto}},
  \bibinfo{journal}{Nat. Photon.} \textbf{\bibinfo{volume}{4}},
  \bibinfo{pages}{367} (\bibinfo{year}{2010}).

\bibitem[{\citenamefont{Cirac et~al.}(1997)\citenamefont{Cirac, Zoller, Kimble,
  and Mabuchi}}]{czkm97}
\bibinfo{author}{\bibfnamefont{J.~I.} \bibnamefont{Cirac}},
  \bibinfo{author}{\bibfnamefont{P.}~\bibnamefont{Zoller}},
  \bibinfo{author}{\bibfnamefont{H.~J.} \bibnamefont{Kimble}},
  \bibnamefont{and} \bibinfo{author}{\bibfnamefont{H.}~\bibnamefont{Mabuchi}},
  \bibinfo{journal}{Phys. Rev. Lett.} \textbf{\bibinfo{volume}{78}},
  \bibinfo{pages}{3221} (\bibinfo{year}{1997}).

\bibitem[{\citenamefont{Yao et~al.}(2005)\citenamefont{Yao, Liu, and
  Sham}}]{yls05}
\bibinfo{author}{\bibfnamefont{W.}~\bibnamefont{Yao}},
  \bibinfo{author}{\bibfnamefont{R.-B.} \bibnamefont{Liu}}, \bibnamefont{and}
  \bibinfo{author}{\bibfnamefont{L.~J.} \bibnamefont{Sham}},
  \bibinfo{journal}{Phys. Rev. Lett.} \textbf{\bibinfo{volume}{95}},
  \bibinfo{pages}{30504} (\bibinfo{year}{2005}).

\bibitem[{\citenamefont{Childress et~al.}(2005)\citenamefont{Childress, Taylor,
  S{\o}rensen, and Lukin}}]{ctsl05}
\bibinfo{author}{\bibfnamefont{L.}~\bibnamefont{Childress}},
  \bibinfo{author}{\bibfnamefont{J.~M.} \bibnamefont{Taylor}},
  \bibinfo{author}{\bibfnamefont{A.~S.} \bibnamefont{S{\o}rensen}},
  \bibnamefont{and} \bibinfo{author}{\bibfnamefont{M.~D.} \bibnamefont{Lukin}},
  \bibinfo{journal}{Phys. Rev. A} \textbf{\bibinfo{volume}{72}},
  \bibinfo{pages}{52330} (\bibinfo{year}{2005}).

\bibitem[{\citenamefont{Waks and Vuckovic}(2006)}]{wv06}
\bibinfo{author}{\bibfnamefont{E.}~\bibnamefont{Waks}} \bibnamefont{and}
  \bibinfo{author}{\bibfnamefont{J.}~\bibnamefont{Vuckovic}},
  \bibinfo{journal}{Phys. Rev. Lett.} \textbf{\bibinfo{volume}{96}},
  \bibinfo{pages}{153601} (\bibinfo{year}{2006}).

\bibitem[{\citenamefont{{T. D. Ladd and P. van Loock and K. Nemoto and W. J.
  Munro and Y. Yamamoto}}(2006)}]{hybridnjp}
\bibinfo{author}{\bibnamefont{{T. D. Ladd and P. van Loock and K. Nemoto and W.
  J. Munro and Y. Yamamoto}}}, \bibinfo{journal}{New J. Phys.}
  \textbf{\bibinfo{volume}{8}}, \bibinfo{pages}{184} (\bibinfo{year}{2006}).

\bibitem[{\citenamefont{Clark et~al.}(2007)\citenamefont{Clark, Fu, Ladd, and
  Yamamoto}}]{susan}
\bibinfo{author}{\bibfnamefont{S.~M.} \bibnamefont{Clark}},
  \bibinfo{author}{\bibfnamefont{K.~M.~C.} \bibnamefont{Fu}},
  \bibinfo{author}{\bibfnamefont{T.~D.} \bibnamefont{Ladd}}, \bibnamefont{and}
  \bibinfo{author}{\bibfnamefont{Y.}~\bibnamefont{Yamamoto}},
  \bibinfo{journal}{Phys. Rev. Lett.} \textbf{\bibinfo{volume}{99}},
  \bibinfo{pages}{40501} (\bibinfo{year}{2007}).

\bibitem[{\citenamefont{Lampert}(1958)}]{Lampert}
\bibinfo{author}{\bibfnamefont{M.}~\bibnamefont{Lampert}},
  \bibinfo{journal}{Phys. Rev. Lett.} \textbf{\bibinfo{volume}{1}},
  \bibinfo{pages}{450} (\bibinfo{year}{1958}).

\bibitem[{\citenamefont{Haynes}(1960)}]{Haynes}
\bibinfo{author}{\bibfnamefont{J.}~\bibnamefont{Haynes}},
  \bibinfo{journal}{Phys. Rev. Lett.} \textbf{\bibinfo{volume}{4}},
  \bibinfo{pages}{361} (\bibinfo{year}{1960}).

\bibitem[{\citenamefont{Thomas and Hopfield}(1961)}]{Hopfield}
\bibinfo{author}{\bibfnamefont{D.}~\bibnamefont{Thomas}} \bibnamefont{and}
  \bibinfo{author}{\bibfnamefont{J.}~\bibnamefont{Hopfield}},
  \bibinfo{journal}{Phys. Rev. Lett.} \textbf{\bibinfo{volume}{7}},
  \bibinfo{pages}{316} (\bibinfo{year}{1961}).

\bibitem[{\citenamefont{Merz et~al.}(1972)\citenamefont{Merz, Kukimoto, Nassau,
  and Shiever}}]{merz}
\bibinfo{author}{\bibfnamefont{J.~L.} \bibnamefont{Merz}},
  \bibinfo{author}{\bibfnamefont{H.}~\bibnamefont{Kukimoto}},
  \bibinfo{author}{\bibfnamefont{K.}~\bibnamefont{Nassau}}, \bibnamefont{and}
  \bibinfo{author}{\bibfnamefont{J.~W.} \bibnamefont{Shiever}},
  \bibinfo{journal}{Phys. Rev. B} \textbf{\bibinfo{volume}{6}},
  \bibinfo{pages}{545} (\bibinfo{year}{1972}).

\bibitem[{\citenamefont{Dean et~al.}(1981)\citenamefont{Dean, Herbert,
  Werkhoven, Fitzpatrick, and Bhargava}}]{dean}
\bibinfo{author}{\bibfnamefont{P.}~\bibnamefont{Dean}},
  \bibinfo{author}{\bibfnamefont{D.}~\bibnamefont{Herbert}},
  \bibinfo{author}{\bibfnamefont{C.}~\bibnamefont{Werkhoven}},
  \bibinfo{author}{\bibfnamefont{B.}~\bibnamefont{Fitzpatrick}},
  \bibnamefont{and} \bibinfo{author}{\bibfnamefont{R.}~\bibnamefont{Bhargava}},
  \bibinfo{journal}{Phys. Rev. B} \textbf{\bibinfo{volume}{23}},
  \bibinfo{pages}{4888} (\bibinfo{year}{1981}).

\bibitem[{\citenamefont{Sanaka et~al.}(2009)\citenamefont{Sanaka, Pawlis, Ladd,
  Lischka, and Yamamoto}}]{kaoru_indistinguishable}
\bibinfo{author}{\bibfnamefont{K.}~\bibnamefont{Sanaka}},
  \bibinfo{author}{\bibfnamefont{A.}~\bibnamefont{Pawlis}},
  \bibinfo{author}{\bibfnamefont{T.~D.} \bibnamefont{Ladd}},
  \bibinfo{author}{\bibfnamefont{K.}~\bibnamefont{Lischka}}, \bibnamefont{and}
  \bibinfo{author}{\bibfnamefont{Y.}~\bibnamefont{Yamamoto}},
  \bibinfo{journal}{Phys. Rev. Lett.} \textbf{\bibinfo{volume}{103}},
  \bibinfo{pages}{053601} (\bibinfo{year}{2009}).

\bibitem[{\citenamefont{Karasyuk et~al.}(1994)\citenamefont{Karasyuk, nd~D.~G.
  S.~Beckett, Nissen, Villemaire, Steiner, and Thewalt}}]{Karasyuk94a}
\bibinfo{author}{\bibfnamefont{V.~A.} \bibnamefont{Karasyuk}},
  \bibinfo{author}{\bibnamefont{nd~D.~G. S.~Beckett}},
  \bibinfo{author}{\bibfnamefont{M.~K.} \bibnamefont{Nissen}},
  \bibinfo{author}{\bibfnamefont{A.}~\bibnamefont{Villemaire}},
  \bibinfo{author}{\bibfnamefont{T.~W.} \bibnamefont{Steiner}},
  \bibnamefont{and} \bibinfo{author}{\bibfnamefont{M.~L.~W.}
  \bibnamefont{Thewalt}}, \bibinfo{journal}{Phys. Rev. B}
  \textbf{\bibinfo{volume}{49}}, \bibinfo{pages}{16381} (\bibinfo{year}{1994}).

\bibitem[{\citenamefont{Tyryshkin et~al.}(2003)\citenamefont{Tyryshkin, Lyon,
  Astashkin, and Raitsimring}}]{Tyryshkin}
\bibinfo{author}{\bibfnamefont{A.}~\bibnamefont{Tyryshkin}},
  \bibinfo{author}{\bibfnamefont{S.}~\bibnamefont{Lyon}},
  \bibinfo{author}{\bibfnamefont{A.}~\bibnamefont{Astashkin}},
  \bibnamefont{and}
  \bibinfo{author}{\bibfnamefont{A.}~\bibnamefont{Raitsimring}},
  \bibinfo{journal}{Phys. Rev. B} \textbf{\bibinfo{volume}{68}},
  \bibinfo{pages}{193207} (\bibinfo{year}{2003}).

\bibitem[{\citenamefont{de~Sousa and Sarma}(2003)}]{desousa}
\bibinfo{author}{\bibfnamefont{R.}~\bibnamefont{de~Sousa}} \bibnamefont{and}
  \bibinfo{author}{\bibfnamefont{S.~D.} \bibnamefont{Sarma}},
  \bibinfo{journal}{Phys. Rev. B} \textbf{\bibinfo{volume}{68}},
  \bibinfo{pages}{115322} (\bibinfo{year}{2003}).

\bibitem[{\citenamefont{Benjamin et~al.}(2006)\citenamefont{Benjamin, Browne,
  Fitzsimmons, and Morton}}]{benjamin_broker}
\bibinfo{author}{\bibfnamefont{S.~C.} \bibnamefont{Benjamin}},
  \bibinfo{author}{\bibfnamefont{D.~E.} \bibnamefont{Browne}},
  \bibinfo{author}{\bibfnamefont{J.}~\bibnamefont{Fitzsimmons}},
  \bibnamefont{and} \bibinfo{author}{\bibfnamefont{J.~J.~L.}
  \bibnamefont{Morton}}, \bibinfo{journal}{New J. Phys.}
  \textbf{\bibinfo{volume}{8}}, \bibinfo{pages}{141} (\bibinfo{year}{2006}).

\bibitem[{\citenamefont{Cakir and Takagahara}(2009)}]{Cakir2009}
\bibinfo{author}{\bibfnamefont{O.}~\bibnamefont{Cakir}} \bibnamefont{and}
  \bibinfo{author}{\bibfnamefont{T.}~\bibnamefont{Takagahara}},
  \bibinfo{journal}{Phys. Rev. B} \textbf{\bibinfo{volume}{80}},
  \bibinfo{pages}{155323} (\bibinfo{year}{2009}).

\bibitem[{\citenamefont{Pawlis et~al.}(2008{\natexlab{a}})\citenamefont{Pawlis,
  Panfilova, As, Lischka, Sanaka, Ladd, and Yamamoto}}]{Pawlis2008}
\bibinfo{author}{\bibfnamefont{A.}~\bibnamefont{Pawlis}},
  \bibinfo{author}{\bibfnamefont{M.}~\bibnamefont{Panfilova}},
  \bibinfo{author}{\bibfnamefont{D.~J.} \bibnamefont{As}},
  \bibinfo{author}{\bibfnamefont{K.}~\bibnamefont{Lischka}},
  \bibinfo{author}{\bibfnamefont{K.}~\bibnamefont{Sanaka}},
  \bibinfo{author}{\bibfnamefont{T.~D.} \bibnamefont{Ladd}}, \bibnamefont{and}
  \bibinfo{author}{\bibfnamefont{Y.}~\bibnamefont{Yamamoto}},
  \bibinfo{journal}{Phys. Rev. B} \textbf{\bibinfo{volume}{77}},
  \bibinfo{pages}{153304} (\bibinfo{year}{2008}{\natexlab{a}}).

\bibitem[{\citenamefont{Panfilova et~al.}(2009)\citenamefont{Panfilova, Pawlis,
  Arens, de~Vasconcellos, Berth, Hüsch, Wiedemeier, Zrenner, and
  Lischka}}]{Panfilova2009}
\bibinfo{author}{\bibfnamefont{M.}~\bibnamefont{Panfilova}},
  \bibinfo{author}{\bibfnamefont{A.}~\bibnamefont{Pawlis}},
  \bibinfo{author}{\bibfnamefont{C.}~\bibnamefont{Arens}},
  \bibinfo{author}{\bibfnamefont{S.~M.} \bibnamefont{de~Vasconcellos}},
  \bibinfo{author}{\bibfnamefont{G.}~\bibnamefont{Berth}},
  \bibinfo{author}{\bibfnamefont{K.~P.} \bibnamefont{Hüsch}},
  \bibinfo{author}{\bibfnamefont{V.}~\bibnamefont{Wiedemeier}},
  \bibinfo{author}{\bibfnamefont{A.}~\bibnamefont{Zrenner}}, \bibnamefont{and}
  \bibinfo{author}{\bibfnamefont{K.}~\bibnamefont{Lischka}},
  \bibinfo{journal}{Microelectronics Journal} \textbf{\bibinfo{volume}{40}},
  \bibinfo{pages}{221} (\bibinfo{year}{2009}).

\bibitem[{\citenamefont{Illing et~al.}(1995)\citenamefont{Illing, Bacher,
  Kummell, Forchel, Anderson, Hommel, Jobst, and Landwehr}}]{Illing}
\bibinfo{author}{\bibfnamefont{M.}~\bibnamefont{Illing}},
  \bibinfo{author}{\bibfnamefont{G.}~\bibnamefont{Bacher}},
  \bibinfo{author}{\bibfnamefont{T.}~\bibnamefont{Kummell}},
  \bibinfo{author}{\bibfnamefont{A.}~\bibnamefont{Forchel}},
  \bibinfo{author}{\bibfnamefont{T.~G.} \bibnamefont{Anderson}},
  \bibinfo{author}{\bibfnamefont{D.}~\bibnamefont{Hommel}},
  \bibinfo{author}{\bibfnamefont{B.}~\bibnamefont{Jobst}}, \bibnamefont{and}
  \bibinfo{author}{\bibfnamefont{G.}~\bibnamefont{Landwehr}},
  \bibinfo{journal}{Appl. Phys. Lett.} \textbf{\bibinfo{volume}{67}},
  \bibinfo{pages}{124} (\bibinfo{year}{1995}).

\bibitem[{\citenamefont{Pawlis et~al.}(2006)\citenamefont{Pawlis, Sanaka,
  G{\"o}tzinger, Yamamoto, and Lischka}}]{Pawlis2006}
\bibinfo{author}{\bibfnamefont{A.}~\bibnamefont{Pawlis}},
  \bibinfo{author}{\bibfnamefont{K.}~\bibnamefont{Sanaka}},
  \bibinfo{author}{\bibfnamefont{S.}~\bibnamefont{G{\"o}tzinger}},
  \bibinfo{author}{\bibfnamefont{Y.}~\bibnamefont{Yamamoto}}, \bibnamefont{and}
  \bibinfo{author}{\bibfnamefont{K.}~\bibnamefont{Lischka}},
  \bibinfo{journal}{Semicond. Sci. Technol.} \textbf{\bibinfo{volume}{21}},
  \bibinfo{pages}{1412} (\bibinfo{year}{2006}).

\bibitem[{\citenamefont{Pawlis et~al.}(2008{\natexlab{b}})\citenamefont{Pawlis,
  Panfilova, As, Lischka, Sanaka, Ladd, and Yamamoto}}]{Kaorulasing}
\bibinfo{author}{\bibfnamefont{A.}~\bibnamefont{Pawlis}},
  \bibinfo{author}{\bibfnamefont{M.}~\bibnamefont{Panfilova}},
  \bibinfo{author}{\bibfnamefont{D.}~\bibnamefont{As}},
  \bibinfo{author}{\bibfnamefont{K.}~\bibnamefont{Lischka}},
  \bibinfo{author}{\bibfnamefont{K.}~\bibnamefont{Sanaka}},
  \bibinfo{author}{\bibfnamefont{T.}~\bibnamefont{Ladd}}, \bibnamefont{and}
  \bibinfo{author}{\bibfnamefont{Y.}~\bibnamefont{Yamamoto}},
  \bibinfo{journal}{Phys. Rev. B} \textbf{\bibinfo{volume}{77}},
  \bibinfo{pages}{153304} (\bibinfo{year}{2008}{\natexlab{b}}).

\bibitem[{\citenamefont{{Hanbury Brown} and Twiss}(1958)}]{HanburyBrown1958}
\bibinfo{author}{\bibfnamefont{R.}~\bibnamefont{{Hanbury Brown}}}
  \bibnamefont{and} \bibinfo{author}{\bibfnamefont{R.~Q.} \bibnamefont{Twiss}},
  \bibinfo{journal}{Proc. R. Soc. Lond. A} \textbf{\bibinfo{volume}{243}},
  \bibinfo{pages}{291} (\bibinfo{year}{1958}).

\bibitem[{\citenamefont{Mandel and Wolf}(1995)}]{mandelwolf}
\bibinfo{author}{\bibfnamefont{L.}~\bibnamefont{Mandel}} \bibnamefont{and}
  \bibinfo{author}{\bibfnamefont{E.}~\bibnamefont{Wolf}},
  \emph{\bibinfo{title}{Optical coherence and quantum optics}}
  (\bibinfo{publisher}{Cambridge University Press}, \bibinfo{year}{1995}).

\bibitem[{\citenamefont{Willmann et~al.}(1973)\citenamefont{Willmann, Dreybodt,
  and Bettini}}]{Willmann}
\bibinfo{author}{\bibfnamefont{F.}~\bibnamefont{Willmann}},
  \bibinfo{author}{\bibfnamefont{W.}~\bibnamefont{Dreybodt}}, \bibnamefont{and}
  \bibinfo{author}{\bibfnamefont{M.}~\bibnamefont{Bettini}},
  \bibinfo{journal}{Phys. Rev. B} \textbf{\bibinfo{volume}{8}},
  \bibinfo{pages}{2891} (\bibinfo{year}{1973}).

\bibitem[{\citenamefont{White et~al.}(1974)\citenamefont{White, Dean, and
  Day}}]{White1974}
\bibinfo{author}{\bibfnamefont{A.}~\bibnamefont{White}},
  \bibinfo{author}{\bibfnamefont{P.}~\bibnamefont{Dean}}, \bibnamefont{and}
  \bibinfo{author}{\bibfnamefont{B.}~\bibnamefont{Day}}, \bibinfo{journal}{J.
  Phys. C: Solid State Phys.} \textbf{\bibinfo{volume}{7}},
  \bibinfo{pages}{1400} (\bibinfo{year}{1974}).

\bibitem[{\citenamefont{Dean et~al.}(1977)\citenamefont{Dean, Bimberg, and
  Mansfield}}]{Dean1977}
\bibinfo{author}{\bibfnamefont{P.}~\bibnamefont{Dean}},
  \bibinfo{author}{\bibfnamefont{D.}~\bibnamefont{Bimberg}}, \bibnamefont{and}
  \bibinfo{author}{\bibfnamefont{F.}~\bibnamefont{Mansfield}},
  \bibinfo{journal}{Phys. Rev. B} \textbf{\bibinfo{volume}{15}},
  \bibinfo{pages}{3906} (\bibinfo{year}{1977}).

\bibitem[{\citenamefont{Wolverson et~al.}(1996)\citenamefont{Wolverson, Boyce,
  Townsley, Schlichterle, and Davies}}]{Wolverson}
\bibinfo{author}{\bibfnamefont{D.}~\bibnamefont{Wolverson}},
  \bibinfo{author}{\bibfnamefont{P.}~\bibnamefont{Boyce}},
  \bibinfo{author}{\bibfnamefont{C.}~\bibnamefont{Townsley}},
  \bibinfo{author}{\bibfnamefont{B.}~\bibnamefont{Schlichterle}},
  \bibnamefont{and} \bibinfo{author}{\bibfnamefont{J.}~\bibnamefont{Davies}},
  \bibinfo{journal}{J. Cryst. Growth} \textbf{\bibinfo{volume}{159}},
  \bibinfo{pages}{229} (\bibinfo{year}{1996}).

\bibitem[{\citenamefont{Duan and Kimble}(2004)}]{duan04}
\bibinfo{author}{\bibfnamefont{L.}~\bibnamefont{Duan}} \bibnamefont{and}
  \bibinfo{author}{\bibfnamefont{H.}~\bibnamefont{Kimble}},
  \bibinfo{journal}{Phys. Rev. Lett.} \textbf{\bibinfo{volume}{92}},
  \bibinfo{pages}{127902} (\bibinfo{year}{2004}).

\end{thebibliography}
\end{document}